\documentclass[usenatbib]{mn2e}
\usepackage{amsfonts}
\usepackage{amsmath}
\usepackage{graphicx}
\usepackage{natbib}

\def\mnras{Mon.\ Not.\ R.\ Astron.\ Soc.}

\def\aap{Astron.\ Astrophys.}
\def\apj{Astrophys.\ J.}

\def\apjs{Astrophys.\ J. Supp.}

\def\prd{Phys.\ Rev.\ D}

\def\physrep{Phys. Rep.}

\title[Non-linear dark matter PDF in redshift space]
      {Ellipsoidal collapse and the redshift space probability 
       distribution function of dark matter}

\author[T. Y. Lam \& R. K. Sheth]
 {Tsz Yan Lam\thanks{E-mail:  tylam@sas.upenn.edu, shethrk@physics.upenn.edu}
  \& Ravi K. Sheth\footnotemark[1]\\
 Department of Physics \& Astronomy, University of Pennsylvania, 
 209 S. 33rd Street, Philadelphia, PA 19104, USA}

\newcommand{\bm}[1]{{\mbox{\boldmath $#1$}}}

\begin{document}
\pagerange{\pageref{firstpage}--\pageref{lastpage}}

\maketitle

\label{firstpage}

\begin{abstract}
We use the physics of ellipsoidal collapse to model the 
probability distribution function of the smoothed dark matter 
density field in real and redshift space.  We provide a simple 
approximation to the exact collapse model which shows clearly 
how the evolution can be thought of as a modification of the 
spherical evolution model as well as of the Zeldovich Approximation 
\citep{zel70}.  
In essence, our model specifies how the initial smoothed overdensity 
and shear fields can be used to determine the shape and size of the 
region at later times.  We use our parametrization to extend previous 
work on the real-space PDF so that it predicts the redshift space 
PDF as well.  Our results are in good agreement with measurements 
of the PDF in simulations of clustering from Gaussian initial 
conditions down to scales on which the rms fluctuation is slightly 
greater than unity.  We also show how the highly non-Gaussian 
non-linear redshifted density field in a numerical simulation can 
be transformed so that it provides an estimate of the shape of the 
initial real-space PDF.  When applied to our simulations, our method 
recovers the initial Gaussian PDF, provided the variance in the 
nonlinear smoothed field is less than 4.
\end{abstract}


\section{Introduction}
The probability distribution function (hereafter PDF; some authors 
prefer to call this the probability density function) specifies the 
probability that a randomly placed cell of specified size and shape 
contains a certain specified density.  As gravitational instability 
alters the spatial distribution of dark matter, the PDF of the dark 
matter density field evolves.  Broadly speaking, there are two 
approaches to modeling this evolution:  one is to study the evolution 
of the lower order moments of the PDF, and the other studies the shape 
of the PDF in its entirety.  

Typically, the first approach attempts to describe the evolution 
exactly, using methods derived from perturbation theory, 
but then the PDF is often computed approximately, upon truncating 
higher order terms in the perturbation theory 
\citep{b94,hept,ptreview}.  
The second uses simple approximations to the exact evolution to 
derive tractable results; in almost all cases, these specify 
local transformations which describe how quantities in the 
initial field determine the evolved density.  
The best studied of these approximations are the spherical 
evolution model \citep{b94,ps97,fg98,gc99,brlc02}, or the Zeldovich 
approximation \citep{br91,pk93,kbg94} and its extension to the 
ellipsoidal collapse model \citep{ecpdf,ls08}.  In principle, 
there is a limit to how well approaches based on deterministic 
transformations can fare \citep{rks98}, but, provided one 
restricts attention to large enough scales, this is a small 
effect \citep{ls08}.  

The associated PDF in redshift, rather than real space, is less 
well-studied.  In this case too, the PDF is described in terms 
of its moments \citep{hbcj95,ptreview}, using the spherical 
model \citep{sg01}, or using the Zeldovich approximation \citep{hks00}.
In addition, \citet{wt01} describe a calculation of the 
PDF based on second order perturbation theory.  
The main goal of the present work is to present a calculation 
of the PDF which is based on the ellipsoidal collapse model.  

Our work is motivated by the fact that the spherical collapse model 
is inconsistent with a linear theory analysis of the second moment 
of the PDF by \citet{kaiser}.  Whereas Kaiser showed that the ratio 
of the redshift space variance to that in real space should be 
 $(1 + 2f/3 + f^2/5)$ (here $f = {\rm d}\ln D/{\rm d}\ln a$ with 
$D$ the linear theory growth factor and $a$ the expansion factor), 
a finding which was confirmed by \citet{fisher95} using a very 
different approach, the spherical model predicts that this factor is 
 $(1 + f/3)^2$ \citep{sg01}.  
An analysis based on the Zeldovich approximation does yield 
Kaiser's factor \citep{ecpdf}, providing yet another, very different 
derivation of this factor, so it is the spherical model which is 
in error.  For this reason, our analysis of the redshift space PDF 
is based on the ellipsoidal collapse model.  Although this model 
has a long history \citep{lms65,vi73,ws79,bs81}, we use it in 
the form given by \citet{bm96}.  They showed how to write this model 
so that, to lowest order it reduces to the Zeldovich approximation.  
It is this formulation we use, so the approach we outline below 
is {\em guaranteed} to correctly reproduce the Kaiser factor. 
This ellipsoidal collapse model (and its associated tidal 
shear effects) has previously been used to motivate more accurate 
models of the abundance of virialized dark matter halos \citep{smt01};
it also provides a framework for studying how the morphology, 
rather than just the density, of the large scale enviroment affect 
structure formation \citep{fitec,d07,ds08}.

We describe our model in Section~\ref{pdfz}, and compare its predictions 
with measurements in a numerical simulation in Section~\ref{section:xpx}. 
Section~\ref{reconst} describes a method for reconstructing the 
shape of the initial real-space PDF from the evolved redshift space PDF.  
A final section summarizes.  Details of some of our methods are 
provided in two Appendices.

\section{The redshift space PDF}\label{pdfz}
In what follows, the variance of the initial density fluctuation field 
when smoothed on scale $R_M$ plays an important role.  
It is denoted by
\begin{equation}
\sigma_{\rm L}^2(M) \equiv \int \frac{{\rm d} k}{k}
\frac{k^3P_{\rm L}(k)}{2\pi^2} |W(kR_M)|^2,
\end{equation}
where $P_{\rm L}(k)$ is the power spectrum of the initial field, 
extrapolated using linear theory to the present time,
$W(x)$ is the Fourier transform of the smoothing window function, and 
$R_M = (3M/4\pi\bar{\rho})^{1/3}$.
In addition, we use $D(t)$ to denote the linear growth factor, and 
$f \equiv {\rm d} \ln D/{\rm d} \ln a \approx \Omega^{5/9}$.

\subsection{The non-linear density in real space}\label{section:rho_r}
The nonlinear overdensity of a region of volume $V$ containing mass $M$ is 
\begin{equation}
 \rho \equiv 1 + \delta = \frac{M}{\bar{\rho}V},
\end{equation}
where $\bar{\rho}$ is the mean density. 
If the evolved Eulerian region is a sphere, then the evolved 
overdensity in real space is well-approximated by 
\begin{equation}
 \rho_r = \frac{(1 - \delta_l/3)^3}{(1-\delta_l/\delta_c)^{\delta_c}}\,
          \prod_{j=1}^3 (1-\lambda_j)^{-1}
\label{eqn:ellreal}
\end{equation}
(see Appendix~\ref{section:EC}).  
Here $\delta_c$ is the critical value of spherical collapse model and 
its exact value depends weakly on the background cosmology. 
In a ${\rm \Lambda}$CDM cosmology $\delta_c \approx 1.66$, while in 
an Einstein-de Sitter model $\delta_c \approx 1.686$.) 
The $\lambda_j$s are the eigenvalues of the deformation tensor, 
extrapolated using linear theory to the present time, 
and $\delta_l \equiv \sum_{j=1}^3\lambda_j$ is the linear theory 
overdensity.  

This approximation has considerable intuitive appeal:
the factor in the product sign is the Zeldovich approximation to 
the evolution, and the factor in front is the ratio of the evolution 
of a sphere in the Zeldovich approximation to that in the spherical 
collapse model.  The expression actually uses a simple approximation 
to the evolution of the density in the spherical model; in principle, 
one could use the exact evolution instead.

\subsection{The non-linear density in redshift space}\label{section:rho_s}
The redshift space overdensity $\rho_s$ is related to the real space 
one by the mapping from real space to redshift space
\begin{equation}
 \vec{s} = \vec{x} + \frac{\vec{v}\cdot\vec{e}}{H}\,\vec{e},
\end{equation}
where $\vec{s}$ is the redshift space coordinate, 
$\vec{x}$ is the real space coordinate, 
$\vec{v}$ is the peculiar velocity at $\vec{x}$, 
$H$ is the Hubble parameter, 
and $\vec{e}$ is the line-of-sight direction unit vector.  Then, 
\begin{equation}
 \rho_s = \rho_r \,
          \left|1 + \frac{1}{H}\frac{\partial v_3}{\partial x_3}\right|^{-1},
\end{equation}
where we assume the line-of-sight direction is along the third axis. 
The velocity of the ellipsoidal region along its principal axis can 
be computed using equation~\eqref{eqn:ECrhosc},
\begin{equation}
 \frac{{\rm d} R_k/{\rm d}t}{HR_k} = 1 -
 \frac{f\left\{R_k^i\lambda_k-A^i_h\delta_l\left[1-(1-\delta_l/\delta_c)^{\delta_c/3-1}\right]/3\right\}}{R_k^i(1-\lambda_k)-A^i_h\left[1-\delta_l/3-(1-\delta_l/\delta_c)^{\delta_c/3}\right]},
\end{equation}
where $f = {\rm d} \ln D/{\rm d} \ln a$ and $D(t)$ is the linear growth factor.
$R_k^i$ is the initial axis length and it is related to $\vec{\lambda}$ by 
equation~\eqref{eqn:revol}. For a randomly oriented ellipsoid we label 
the Euler Angle components 
 $(e_1,e_2,e_3)=(\cos\psi\sin\theta,\sin\psi\sin\theta,\cos\theta)$ 
for rotation from the line-of-sight coordinate to the coordinate of 
the principal axes of the ellipsoid. 
As a result the redshift space overdensity is 
\begin{equation}
\begin{split}
 \rho_s= & \frac{(1 - \delta_l/3)^3}{(1-\delta_l/\delta_c)^{\delta_c}} \left(\prod_{j=1}^3 \frac{1}{1-\lambda_j}\right) \times\\
& \left[1-\sum_{k=1}^3\frac{f\left\{R_k^i\lambda_k-A^i_h\delta_l\left[1-(1-\delta_l/\delta_c)^{\delta_c/3-1}\right]/3\right\}}{R_k^i(1-\lambda_k)-A^i_h\left[1-\delta_l/3-(1-\delta_l/\delta_c)^{\delta_c/3}\right]}e_k^2 \right]^{-1}.
\end{split}
\label{eqn:ellred}
\end{equation}
Notice that $\rho_s\to\rho_r$ when $f\to 0$, as it should.

\subsection{The PDF from ellipsoidal collapse} \label{section:PT}
Our redshift space perturbation theory calculation makes use of 
the same assumptions as in real space: there is a local mapping from 
the eigenvalues $\lambda_i$ of the deformation tensor to the non-linear 
overdensity $\rho_s$; and the smoothing scale associated with $\rho_s$ in 
the initial field is the one containing the same mass with the correct 
shape.  
The most important difference in redshift space calculation is that 
the mapping from $\lambda_i$ to $\rho_s$ is no longer deterministic: 
the orientation of the ellipsoid with respect to the line of sight 
introduces stochasticity which must be integrated over.  
As a result, the evolved (Eulerian) PDF at fixed $V$ is related to 
the initial (Lagrangian) PDF at fixed mass scale $M$ by 
\begin{equation}
 \rho_s\, p(\rho_s|V){\rm d}\rho_s = \int {\rm d}{\bm \lambda}\,
     {\rm d}{\bm e}\,p({\bm \lambda}|\sigma)\,
      \delta_{\rm D}\left[\rho_s = \rho_s'({\bm \lambda},{\bm e})\right],
 \label{eqn:ecpdf}
\end{equation}
where $\rho_s \equiv M/\bar{M}$ and $\rho_s'({\bm \lambda},{\bm e})$ is 
the redshift space overdensity given in equation~(\ref{eqn:ellred}).  

In practice the joint distribution of $p({\bm \lambda}|\sigma)$ is a 
function of $\lambda_i/\sigma$ \citep{grf}, where 
\begin{equation}
 \sigma = \sigma_{\rm L}^{\rm sph}(M) \,
  \exp\left\{-\frac{B}{2}\sum_{k<j}\left[\ln\left(\frac{R_j^i}{R_k^i}\right) 
       \right]^2\right\} .
 \label{eqn:ellvar}
\end{equation}
Here $R_j^i$ denotes the initial size of the $j$th principal axis, and
$B=0.0486$ \citep{brlc02}.  
As a result, it is useful to rewrite equation~(\ref{eqn:ecpdf}) as
\begin{equation}
\rho_s\, p(\rho_s|V)\,{\rm d}\rho_s  = 
 \int {\rm d}{\bm \lambda}\,{\rm d}{\bm e}\,p({\bm \lambda}|1) \,
 \delta_{\rm D}\left[\rho_s = \left|\rho_s'({\sigma\bm \lambda},{\bm e})\right|
    \right].
\end{equation}

\begin{figure*}
\centering
\includegraphics[width=0.9\linewidth]{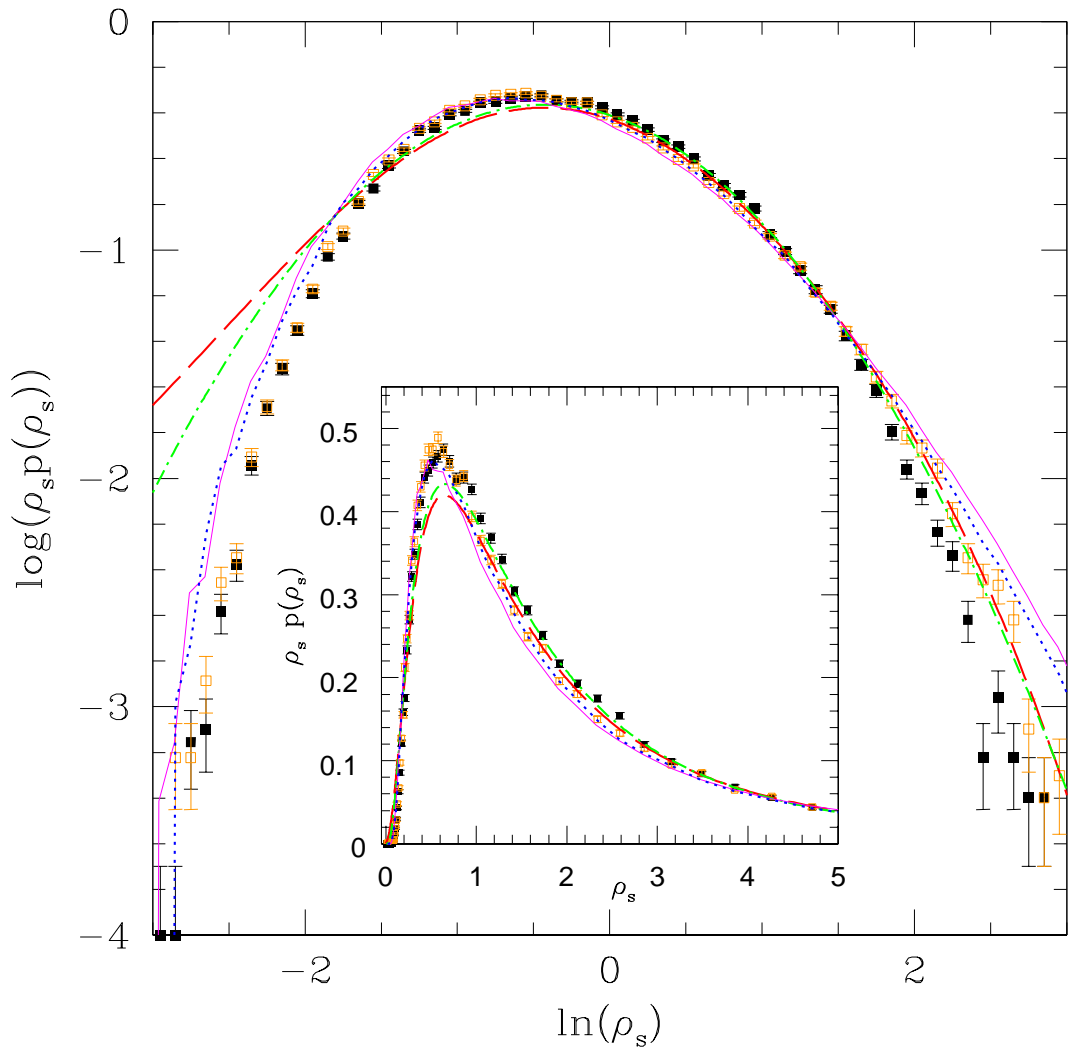}
\caption{Comparions of the measured PDFs with various models in  
         $8h^{-1}{\rm Mpc}$ sphere. 
         The outer panel shows the log-ln plot and the inner panel shows 
         the linear plot.
         Filled (black) squares are measurements from the simulation, 
         open (orange) squares are measurements with finger-of-god removed. 
         Dotted (blue) curve is the prediction of the perturbation theory 
         by applying equation~\eqref{eqn:ellred}. 
         Solid (magenta) curve is the prediction by setting $B=0$ in 
         equation~\eqref{eqn:ellvar}.
         Dot-dashed (green) curve and long-dashed (red) curve are 
         the lognormal and real-redshift space mapping empirial models 
         respectively.
         }
\label{fig:xpxr8}
\end{figure*}

\begin{figure*}
\centering
\includegraphics[width=0.45\linewidth]{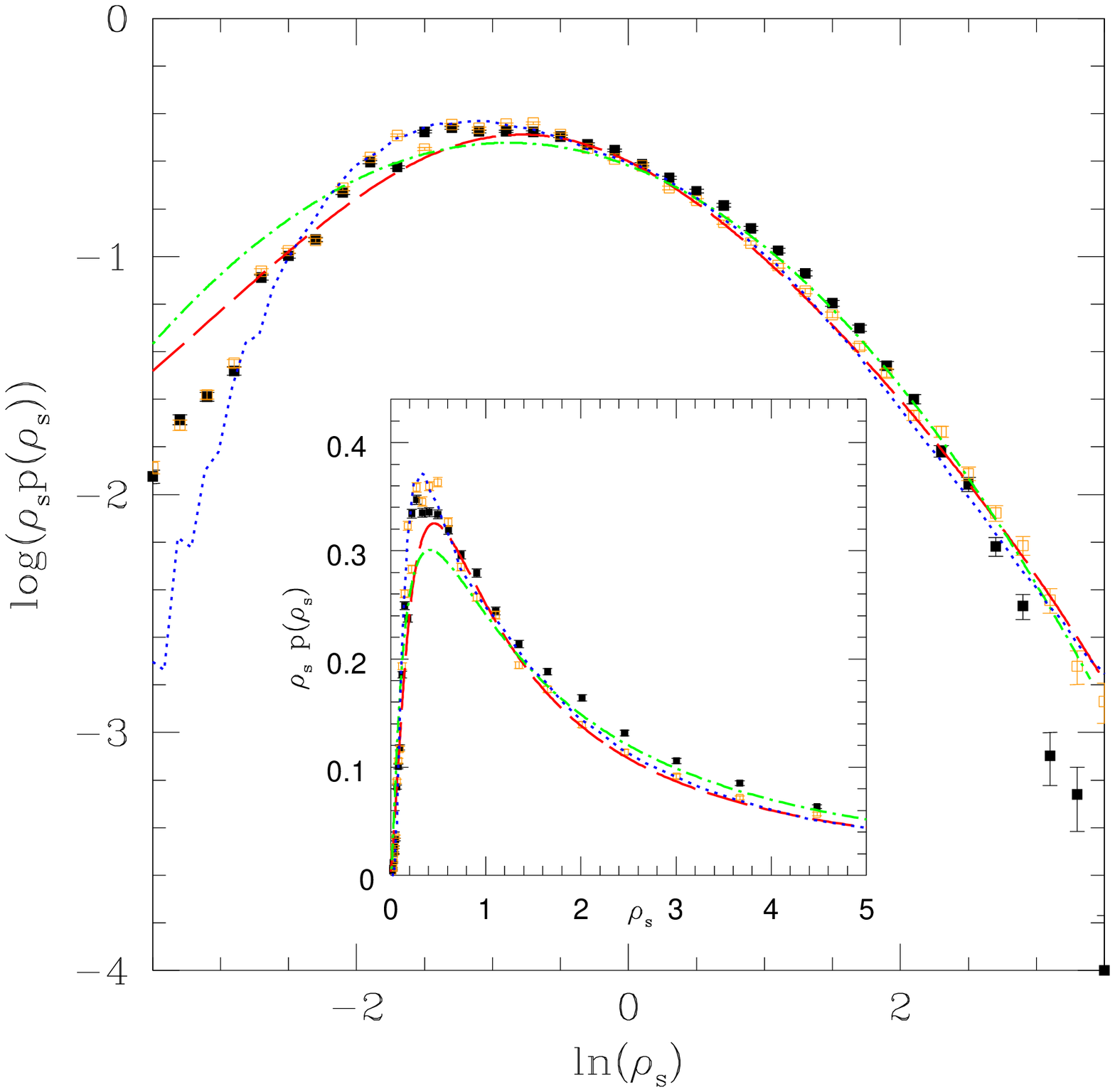}
\includegraphics[width=0.45\linewidth]{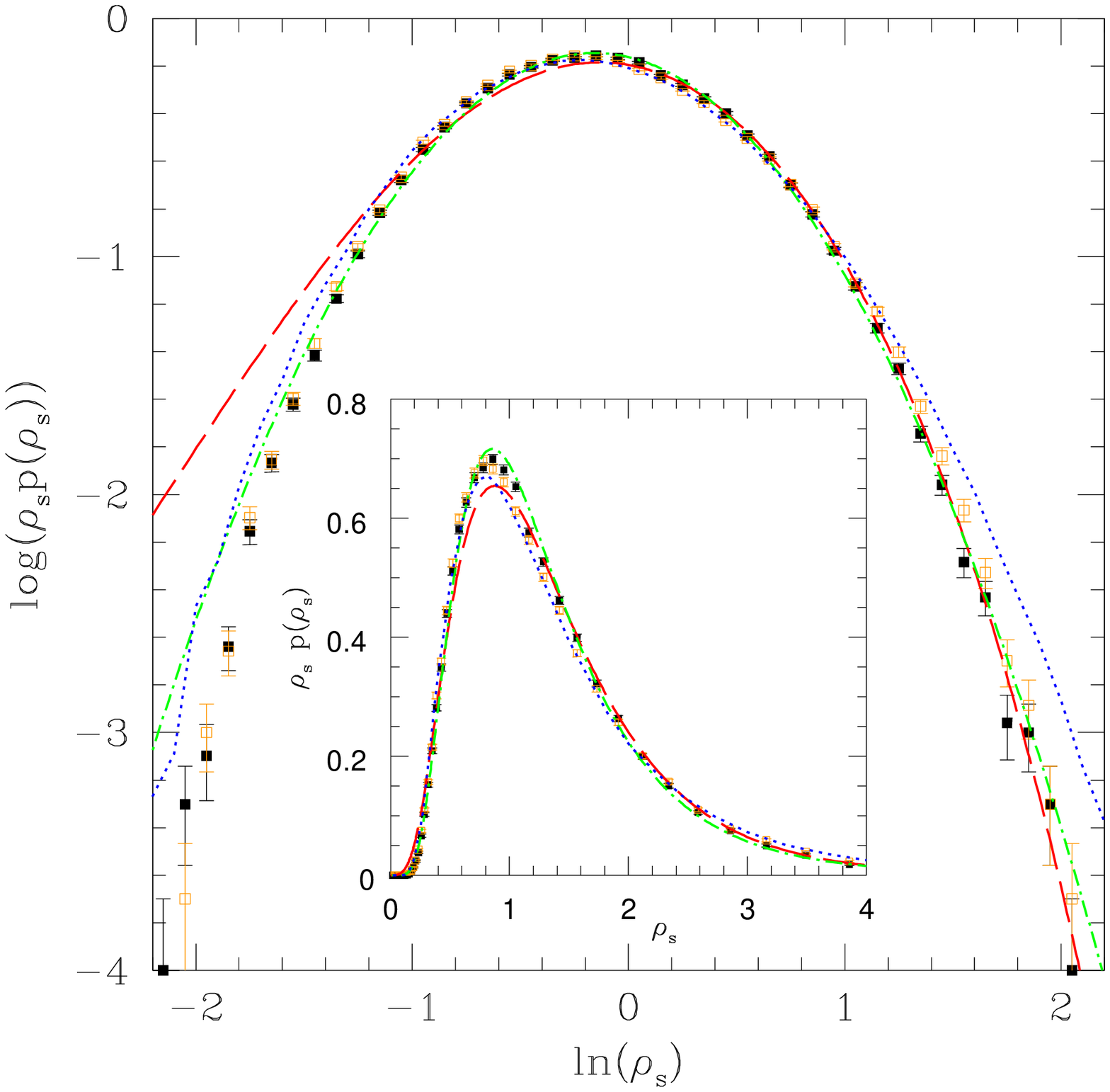}
\caption{Same as previous figure~\ref{fig:xpxr8} but 
         in spheres of radius 4$h^{-1}{\rm Mpc}$ (left) and 
         16$h^{-1}{\rm Mpc}$ (right). 
         The prediction with spherical variance (the solid (magenta) curve 
         in figure~\ref{fig:xpxr8}) is not shown.}
\label{fig:xpxr4+16}
\end{figure*}

As happens for the real space calculation \citep{kbg94,sg01,ecpdf}, 
this estimate of the redshift space PDF is not correctly normalized:
\begin{equation}
 N \equiv \int^{\infty}_0 {\rm d}\rho_s\,p(\rho_s) \ne 1.
 \label{normalize}
\end{equation}
(On the other hand, the integral of equation~\ref{eqn:ecpdf} over 
all $\rho_s$ does equal unity.) 
Therefore, we set 
\begin{equation}
 \rho_s' \equiv N\rho_s, \qquad \text{\rm and}\qquad
 \rho_s'^2\, p'(\rho_s') = \rho_s^2\, p(\rho_s).
\end{equation}
This ensures that $\int {\rm d}\rho_s'\,\rho_s'\,p'(\rho_s')$ 
and $\int {\rm d}\rho_s'\,p'(\rho_s')$ both equal unity
\citep{ls08}.  
In practice, for the scales of interest in what follows, 
(cells of radius $4$, $8$, and $16h^{-1}{\rm Mpc}$), 
$N$ is only slightly larger than unity:  1.264, 1.104 and 1.052.

\subsection{Other approximations}
We will compare our model for the PDF with two other useful 
approximations which are less physically-motivated.  
One is the Lognormal, 
\begin{equation}
 \rho_s p(\rho_s) = \frac{\exp[-(\ln\rho_s + \mu_s)^2/2\sigma_s^2]}
                    {\sqrt{2\pi}\sigma_s},
 \label{eqn:lognormal}
\end{equation}
where $\mu_s= \sigma_s^2/2$ and $\sigma_s^2 = \ln\langle\rho_s^2\rangle$.  
and $\sigma_s^2$ is given by multiplying the Kaiser factor by 
the \citet{smithetal} fitting formula for the real space variance 
$\sigma_r^2$.

The second is based on the finding that, in simulations, the 
distribution of $(\delta_s/\sigma_s)$ is almost the same as 
that of $(\delta_r/\sigma_r)$ \citep{sg01}.  Since the spherical 
model provides a good description of the $(\delta_r/\sigma_r)$ 
in real space \citep{ls08}, we use it together with the Kaiser 
factor to set 
\begin{equation}
 p(\rho_s) = \frac{\sigma_r}{\sigma_s}\,p_{\rm sc}(\rho_r),
 \label{eqn:SCmap}
\end{equation}
where $p_{\rm sc}(\rho)$ is the real space PDF with the spherical 
collapse model.  In essence, this can be thought of as a model 
in which the redshift-space overdensity is a linearly biased 
version of the real-space overdensity, the constant of 
proportionality (the bias factor) being $(\sigma_s/\sigma_r)$.  
\citet{kaiser} suggests that this should be an excellent 
approximation on large scales; although \citet{rs04} suggests 
that it is not a well-motivated approximation on intermediate 
or small scales, measurements of the redshift space PDF by 
\citet{sg01} suggest that it is nevertheless reasonably accurate.

\section{Comparison with simulation} \label{section:xpx}
We compare our models with measurements in a simulation which was 
run and analyzed by \citet{sss07}.
The simulation box is a periodic cube $512 h^{-1}{\rm Mpc}$ on 
a side in a cosmological model where 
 $(\Omega_{\rm m},\Omega_{\rm \Lambda},h,\sigma_8)=(0.27,0.73,0.72,0.9)$. 
The simulation followed the evolution of $400^3$ particles, 
each of mass $1.57 \times 10^{11} h^{-1}{\rm M_{\odot}}$.
Halos were identified using the FOF algorithm with link length 0.2 
times the mean interparticle separation.  

To measure the Eulerian real-space PDFs we place spheres of volume 
$V$ randomly in the simulation box, and record the number of 
particles in each sphere, taking care to account for periodic 
boundary conditions.  We measure the redshift-space PDFs similarly, 
after transforming the particle positions as follows.  
We assume that one of the sides of the box lies parallel to the 
line of sight.  The redshift space coordinate is then given by 
adding the peculiar velocity (divided by the Hubble constant) in 
this direction to the real-space coordinate, taking care to account 
for the periodicity of the box.  

Technically, our model of the PDF is for a continuous density field 
rather than for discrete particle counts.  In practice, we are 
interested in large enough volumes that discreteness effects are 
negligible, so we simply set
 $\rho \equiv 1 + \delta = N/\bar{N}$.  
However, there is another effect which is more pernicious.  
The `virial' motions of particles within halos give rise to long 
fingers-of-god --- this makes halos stretch about 7 times longer 
along the line of sight than across --- an effect first noted 
by \citet{jj72}, which is entirely absent in the perturbation 
theory calculation of the redshift space PDF.  
To illustrate their effect on the PDF, we have also computed 
redshift space positions after first setting all virial motions to 
zero, so that dark matter particles have the same speed as their 
associated halos.  (I.e. we assign all particles in a halo the 
same velocity as that of the halo center of mass.)
In the figures which follow, filled black squares show the 
true redshift space measurements, and open orange squares show 
results when fingers-of-god are absent.  

Figure~\ref{fig:xpxr8} shows the redshift space PDF of the dark 
matter in spheres of radius $8h^{-1}{\rm Mpc}$.  The main panel 
shows this using a $\log$-$\ln$ scale to emphasize the behaviour of 
the low probability tails, and the inset shows the same data using 
linear axes to highlight the behaviour near the peak of the distribution.
Filled and open squares show the measurements in the simulation 
(i.e., before and after compressing the fingers-of-god).  These 
indicate that fingers-of-god are most important in the denser cells; 
this is not unexpected, since dense cells host the most massive halos 
which have the largest virial motions. 

Our perturbation theory based prediction is represented by the 
dotted (blue) curve --- notice that it provides a significantly 
better fit to the simulations when the fingers-of-god have been 
compressed.  
The solid (magenta) curve is the perturbation theory based prediction 
with $B=0$ in equation~\eqref{eqn:ellvar}, which corresponds to 
using our ellipsoidal collapse model but ignoring the effect 
on the variance associated with differences from a spherical shape. 
The dotted curve ($B=0.0486$) is always closer to the 
measurements than is the solid one ($B=0$), consistent with the 
real-space results presented in \citet{ls08}), although the 
difference between the two curves is not large.  
In what follows, we only show results for the case when $B=0.0486$.  
The dot-dashed (green) curve shows the Lognormal 
(equation~\ref{eqn:lognormal}) and the long-dashed (red) 
curve shows equation~\eqref{eqn:SCmap}.  Our model provides 
a better fit than these other more empirical curves, except in 
the highest density tail.  

Figure~\ref{fig:xpxr4+16} shows similar results for spheres of 
radius 4$h^{-1} {\rm Mpc}$ and 16$h^{-1} {\rm Mpc}$.
In the larger cells, fingers-of-god are irrelevant, and the 
Lognormal provides a slightly better description than does 
our model, which again overshoots at large densities.  
The approach which assumes a simple mapping from real to redshift 
space overdensity fairs reasonably well at large densities, but 
underestimates the height of the peak of the distribution, and 
overestimates the PDF at lower densities.  Our model is 
substantially better than the other two in the smaller 
4$h^{-1} {\rm Mpc}$ cells.

\section{Reconstruction of the initial distributions}\label{reconst}
In \citet{ls08} we described how the spherical collapse 
model could be used as the basis of a method for reconstructing the 
shape of the initial PDF from the Eulerian one.  In principle, the 
method can be used as a consistency check for Gaussian initial 
conditions.  We showed that, in real space, it can reconstruct the 
Gaussian form of the initial distribution from the highly skewed 
nonlinear PDF on scales where the rms fluctuation is about $2$ or less. 

Extending this reconstruction method to redshift space is 
complicated by the fact that we know the spherical model is 
inadequate (it does not reproduce the Kaiser factor).  However, 
the ellipsoidal collapse model, which we have shown works quite well, 
has many more parameters --- the lengths and orientations of the 
three principal axes --- so that the final density is determined 
by a stochastic rather than deterministic mapping.  

So we have tried a simpler approach which is based on the fact 
that equation~(\ref{eqn:SCmap}) provides a reasonable description 
of the redshift space PDF.  Namely, we assume that
\begin{equation}
 \delta_s = \delta_r\,(\sigma_s/\sigma_r), 
 \label{linearbiasz}
\end{equation}
and use the (square root of the) Kaiser factor in place of 
$\sigma_s/\sigma_r$; we have found that using our ellipsoidal 
collapse calculation of this quantity instead produces slightly 
worse results.  Then we set 
\begin{equation}
 \nu \equiv \frac{1 - [(1+\delta_{\rm s}\,\sigma_r/\sigma_s)
              /N_{\rm sc}]^{-1/\delta_{\rm c}}}
            {\sigma_{\rm L}(M/N_{\rm sc})/\delta_{\rm c}},
 \label{rescalenu}
\end{equation}
where $\delta_{\rm c}$ is the critical linear density associated 
with collapse in the spherical model, and $N_{\rm sc}$ is the 
normalization factor (the analogue of equation~\ref{normalize}) 
in the spherical model calculation of the real space PDF. 
Finally, we make a histogram of $\nu$ by weighting each cell 
which contributes to $p(\delta_s)$ by its value of 
$(1+\delta_s\sigma_r/\sigma_s)/N_{\rm sc}$.

\begin{figure}
\centering
\includegraphics[width=\linewidth]{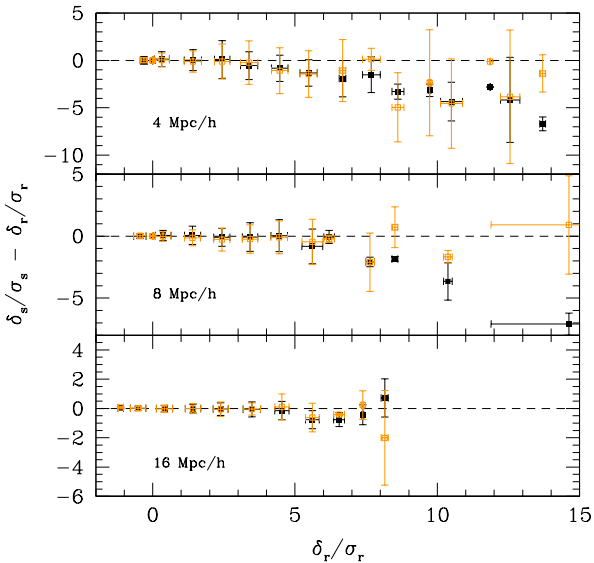}
\caption{Comparsion of real- and redshift-space overdensities in 
         cells of radius $4h^{-1}$Mpc (top), $8h^{-1}$Mpc (middle) 
         and $16h^{-1}$Mpc (bottom).  Filled and open symbols are 
         before and after fingers of god have been removed.  
         They show the mean, and error bars show the rms, for narrow 
         bins in $(\delta/\sigma)_r$.
         If equation~(\ref{linearbiasz}) is accurate, then 
         $(\delta/\sigma)_s - (\delta/\sigma)_r = 0$ with 
         little scatter.  }
\label{fig:nurs}
\end{figure}

Before showing how well this works, Figure~\ref{fig:nurs} shows a 
test of our assumption that equation~(\ref{linearbiasz}) is accurate, 
meaning that $\delta_s$ is linearly related to $\delta_r$, with 
little scatter around the mean relation.  Evidently, these 
assumptions are reasonable on large scales (bottom and middle 
panels), but become increasingly worse on smaller scales (top).  

\begin{figure}
\centering
\includegraphics[width=\linewidth]{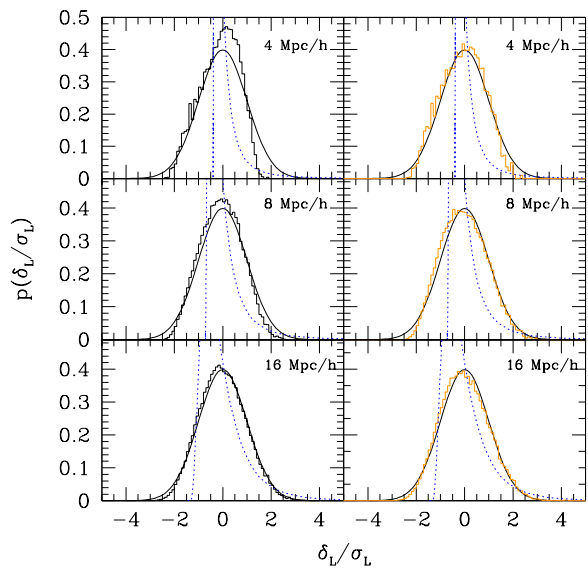}
\caption{Comparsion of the reconstructed linear PDF of $\delta/\sigma$ 
         (histogram) with the expected zero-mean unit-rms Gaussian form 
         (solid curve). Upper panels show results of reconstruction for 
          $4h^{-1} {\rm Mpc}$ scale, 
          middle panels show results for $8h^{-1}{\rm Mpc}$. 
          lower panels show results for $16h^{-1}{\rm Mpc}$. 
          Left- and right-hand panels show results of reconstructions 
          which begin from the filled and open squares in the previous 
          Figures (i.e. estimates of $p(\delta_s)$ with and without 
          fingers-of-god).  Dotted curves in each panel show the 
          corresponding nonlinear PDFs (equation~\ref{eqn:ecpdf}).}
\label{fig:linearmap}
\end{figure}

Figure~\ref{fig:linearmap} shows the reconstruction results for 
$4$, $8$ and $16 h^{-1}{\rm Mpc}$ spheres. 
The left- and right-hand panels show the result of transforming the 
PDFs shown by filled and open squares in the previous Figures 
(i.e., full and compressed fingers of god).  
Smooth solid curves show a Gaussian distribution for comparison. 
The reconstruction method works well for the $8$ and 
$16 h^{-1}{\rm Mpc}$ scales.  This method even works for the 
smaller $4 h^{-1}{\rm Mpc}$ scales if the fingers-of-god have 
been removed. The dotted curves show the corresponding nonlinear 
distributions (equation~\ref{eqn:ecpdf}) to demonstrate that 
our reconstruction method transforms significantly skewed 
distributions back to the original symmetric gaussian shape.

\section{Discussion}\label{section:discussion}
We have extended our previous work on the real space dark matter PDF 
to redshift space.  To do so, we provided a simple parametrization of 
the evolution of the nonlinear density in the ellipsoidal collapse 
model.  This shows explicitly how the model extends both the 
spherical collapse model, and the Zeldovich approximation 
(equations~\ref{eqn:ECrhosc} and \ref{eqn:ellred}).   
This parametrization, with our prescription for relating spatial 
statistics in the initial and evolved fields (equation~\ref{eqn:ecpdf}), 
results in excellent agreement with measurements from a numerical 
simulation, in both real (Figure~\ref{fig:realPDF}) and 
redshift space (Figures~\ref{fig:xpxr8} and~\ref{fig:xpxr4+16}). 
Our parametrization fares significantly better than analyses which 
are based on exact second and third order perturbative expansions 
of the ellipsoidal collapse evolution --- these produce PDFs with 
a power-law tail at high densities which is a signature that the 
analysis is breaking down --- a similar tail was seen for the 
Zeldovich approximation \citep{hks00}.  
Our approach also fares better than the Lognormal model, and 
one which assumes that redshift space overdensities are linearly 
proportional to those in real-space:  
 $\delta_s = (\sigma_s/\sigma_r)\, \delta_r$ 
where the constant of proportionality is given by the (square-root 
of the) ratio of the variances.  The latter is known to be a good 
approximation on large scales --- we find that it is accurate on 
small scales, except in underdense regions.  

We also used our results to motivate a method for reconstructing the 
shape of the initial PDF from the nonlinear evolved redshift space 
PDF.  This method works well on scales where the rms fluctuation is 
less than 2 when fingers-of-god have been removed 
(Figure~\ref{fig:linearmap}).  
We are in the process of applying this method to reconstruct the 
baryonic ascoustic oscillation (BAO) peak in the correlation function 
\citep{esss07}. 
Instead of using equation~(\ref{rescalenu}) it is also possible to 
formulate the reconstruction using equations~\eqref{eqn:ECrhosc} 
or \eqref{eqn:ellred}; this is also the subject of work in progress.

While this provides a nice graphic reconstruction of the 
Gaussianity of the initial conditions, more work is necessary 
before it can be used to distinguish between a truly Gaussian model 
and the very mildly non-Gaussian models which are not yet excluded 
by the CMB \citep{cmb5yr}.  In this respect, it may be interesting 
to extend our method for estimating the nonlinear PDF to the family 
of non-Gaussian models which are currently studied in the context 
of using cluster abundances and clustering to constrain primordial 
non-Gaussianity \citep{shshp08,fnlhalo,fnlverde}.

\section*{Acknowledgements}
We thank
 R. Scherrer for pointing out that there was little understanding of 
 why, when scaled by their variances, the real and redshift space PDFs 
 were so similar, 
 L. Hui for urging us to go beyond an order-by-order analysis, 
 R. E. Smith for the simulation data, 
 and the referee, V. Desjacques, for a thoughtful report.  
This work was supported in part by NSF Grant AST-0507501.

\appendix
\section{Relation between initial and final density in the 
         ellipsoidal collapse model}
\label{section:EC}
In this Appendix we derive relatively simple analytic approximations 
to the evolution of the density within a triaxial perturbation.  
The exact evolution of the perturbation is assumed to be described 
by the ellipsoidal collapse model of \citet{bm96}.
\citet{fitec} provide a simple but useful approximation to 
this evolution which is based on a beautiful analysis of the 
collapsing ellipsoid problem by \citet{ws79}.
It is this approximate solution which we exploit.  
  
\subsection{The collapse model}
Gravity evolves the initial axes $(R_1^i,R_2^i,R_3^i)$ of a 
triaxial object to $(R_1,R_2,R_3)$ because 
\begin{equation}
 \frac{{\rm d}^2 R_k}{{\rm d} t^2} = H_0^2\Omega_{\Lambda}R_k 
   - 4\pi G\bar{\rho}R_k\,
  \left(\frac{1+\delta}{3}+\frac{b_k'}{2}\delta+\lambda'_{{\rm ext,}k}\right),
\label{eqn:ECDE}
\end{equation}
where the interior and exterior tidal forces are respectively,
\begin{align}
b'_k(t)  = &\left(\prod_{i=1}^3R_i(t)\right) \int_0^{\infty}\frac{{\rm d}\tau}
{[R_k^2(t)+\tau]\prod_{j=1}^3[R_j^2(t)+\tau]^{1/2}}  \nonumber \\
        &  - \frac{2}{3}, \\
\lambda'_{{\rm ext,}k}(t)  = & \frac{D(t)}{D(t_i)}\left[\lambda_k(t_i)-\frac{\delta_i}{3}\right]
\end{align}
\citep{bm96}.  
Here $\lambda_k$ are the initial eigenvalues of the deformation tensor, 
and the initial conditions of equation~(\ref{eqn:ECDE}) are set by the 
Zeldovich Approximation:
\begin{align}
 R_k(t_i) & =  a_i[1-\lambda_k(t_i)],\\
 \frac{{\rm d} R_k(t_i)}{{\rm d} t} & =  H(t_i)R_k(t_i)-a(t_i)H(t_i)f_1
 \lambda(t_i),
\end{align}
where $f_1 = {\rm d}\ln D/{\rm d} \ln a$ and $D(t)$ is the linear 
growth factor.

Although equation~(\ref{eqn:ECDE}) in general must be solved numerically, 
\citet{fitec} showed that the following approximation was rather 
accurate:
\begin{equation}
\begin{split}
R_k(t) = & \frac{a(t)}{a_i}R_k(t_i)[1-D(t)\lambda_k]\\
 & -  \frac{a(t)}{a_i}R_h(t_i)\left[1 - \frac{D(t)}{3}\delta_i- 
                                     \frac{a_e(t)}{a(t)}\right],
\end{split}
\label{eqn:fitEC}
\end{equation}
where $R_h(t_i) = 3/\sum_jR_j(t_i)^{-1}$ and $a_e(t)$ is the expansion 
factor of a spherical universe with initial overdensity 
$\delta_i = \sum_j \lambda_j(t_i)$.  This expression is inspired by 
an analysis of collapsing ellipsoids by \citet{ws79}.   
The first term in the expression above is the Zeldovich Approximation 
to the evolution; the second term is a correction to this approximation.  
When $\lambda_2=\delta_i/3$ then the evolution of the second axis is 
almost exactly like that of a sphere with initial overdensity $\delta_i$.  

\subsection{Evolution to a spherical volume}
For an Eulerian sphere of $R_{\rm E}$, equation~(\ref{eqn:fitEC}) 
implies that 
\begin{equation}
R_1^i[1 - D(t)\lambda_1] = R_2^i[1-D(t)\lambda_2] = R_3^i[1-D(t)\lambda_3].
\end{equation}
Substitute into the definition of $R_h(t_i)$
\begin{equation}
R_h(t_i) = \frac{3}{\sum_j R_j(t_i)^{-1}} 
         = \frac{R_k(t_i)(1-D(t)\lambda_k)}{1-D(t)\delta_i/3},
\end{equation}
where $k=1,2,3$.
Therefore we can write the evolved $R_k(t)$ as
\begin{equation}
 \frac{R_{\rm E}}{R_k^i} = 
  \frac{a_e(t)/a(t)}{1 - D(t)\delta_i/3}[1-D(t)\lambda_k],
\label{eqn:revol}
\end{equation}
where $R_k^i = R_k(t_i)$.  This writes the evolution as that in 
the Zeldovich Approximation, times a correction factor which is 
simply the ratio of the evolved and initial radii of a Zeldovich 
sphere $(1 - D\delta_i/3)$ divided by the same ratio in the 
exact spherical evolution model.  Thus, the evolved overdensity is 
\begin{align}
 \rho & = 1+\delta  = \prod_{j=1}^3 \frac{R_j^i}{R_{\rm E}} 
       \approx \frac{(1- D(t)\delta_i/3)^3}{(a_e(t)/a(t))^3}
               \, \prod_{j=1}^3 \frac{1}{1-D(t)\lambda_j} \nonumber\\
      & \approx 
         \frac{(1- D(t)\delta_i/3)^3}{(1 - D(t)\delta_i/\delta_c)^{\delta_c}}
          \, \prod_{j=1}^3 \frac{1}{1-D(t)\lambda_j}.
 \label{eqn:ECrhosc}
\end{align}
The final expression uses a simple approximation for the evolution in 
the spherical collapse model:
\begin{equation}
 \frac{a(t)}{a_e(t)} \approx 
 \left(1 - D(t)\frac{\delta_i}{\delta_c}\right)^{-\delta_c/3}
\label{eqn:ae}
\end{equation}
\citep{b94,rks98}, where $\delta_c\approx 5/3$.  
Notice that when all the eigenvalues are identical, then 
equation~(\ref{eqn:ECrhosc}) reduces to the spherical collapse 
model. 

\begin{figure}
\centering
\includegraphics[width=\linewidth]{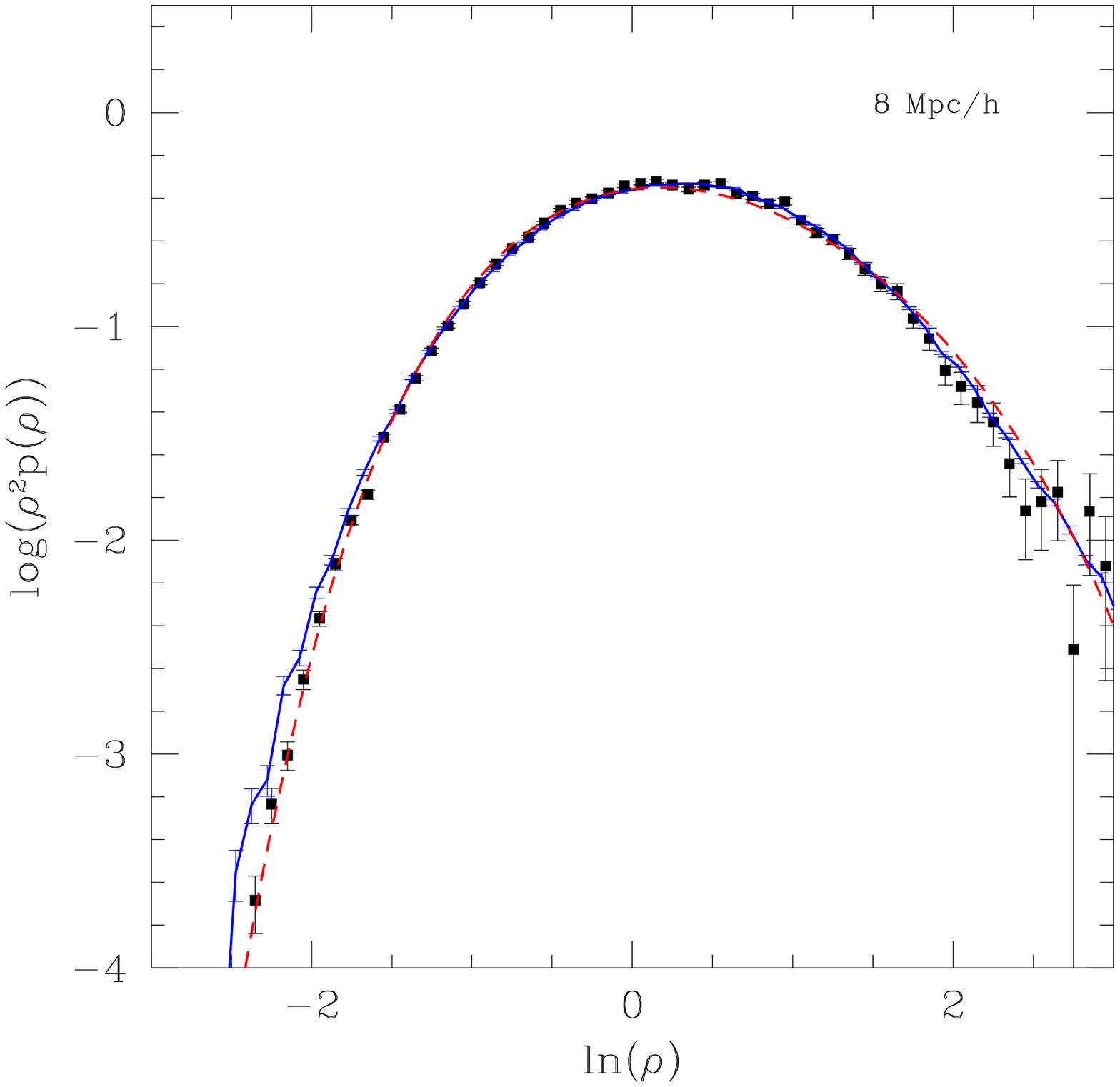}
\caption{The real space non-linear PDF.  Symbols show measurements 
         in a numerical simulation (from Lam \& Sheth 2008). 
         The solid (blue) curve uses our approximation to ellipsoidal 
         collapse (equation~\ref{eqn:ECrhosc}) and the dashed (red) 
         curve is based on our approximation to spherical collapse 
         (equation~\ref{eqn:ae}).}
\label{fig:realPDF}
\end{figure}

Insertion of equation~\eqref{eqn:ae} or~\eqref{eqn:ECrhosc} into 
equation~\eqref{eqn:ecpdf} of the main text yields models for the 
redshift-space PDF associated with spherical and ellipsoidal collapse, 
respectively.  The corresponding models for the real-space PDF are 
given by setting $f=0$ in that expression.  These real-space models 
are shown by the dashed (red) or solid (blue) curves; both are in 
excellent agreement with the measured real-space PDF (symbols).

\subsection{Evolution to an arbitrary shape}
When the Eulerian volume in consideration is triaxial, the computation of 
the overdensity is less trivial. In general one must solve a system of 
non-linear equations for $x_k \equiv a(t)R_k^i/a_iR_k(t)$ in 
\begin{equation}
\begin{split}
 x_k[1 - & D(t)\lambda_k]  =  \\
  & 1 + 3\left(1- \frac{D(t)\delta_i}{3} - \frac{a_e(t)}{a(t)}\right)
         \left( \sum_{j=1}^3\frac{R_k}{R_j}\frac{1}{x_j}\right)^{-1}.
\end{split}
\end{equation}
The overdensity is $1+\delta = \prod_{j=1}^3 x_j$. 
The above equation can be solved numerically or iteratively with 
\begin{align}
 x_k^{(n+1)}= & 
  \left[1 + 3\left(1- \frac{D(t)\delta_i}{3} - \frac{a_e(t)}{a(t)}\right)
   \left( \sum_{j=1}^3\frac{R_k}{R_j}\frac{1}{x_j^{(n)}}\right)^{-1}\right] 
\nonumber \\
 & \times [1 -  D(t)\lambda_k]^{-1} \\
 x_k^{(0)}  = & \frac{1- D(t)\delta_i/3}{a_e(t)/a(t)} 
                \frac{1}{1-D(t)\lambda_k}.
\end{align}
An analytical expression for the overdensity of the ellipsoid can be 
obtained if we make the approximation that the collapse of the second 
axis is very close to that in the spherical model.  \citet{fitec} 
show why this is almost always an excellent approximation.  
One can then approximate the collapse of the second axis using either 
of the two expressions below:
\begin{equation}
\frac{R_2^i}{R_2(t)} \approx
\left\{\begin{aligned}
 &  \frac{1-D(t)\delta_i/3}{a_e(t)/a(t)}\frac{1}{1-D(t)\lambda_2}, \\
 & \left(1 - \frac{D(t)\delta_i}{\delta_c}\right)^{-\delta_c/3}.
\end{aligned}\right.
\end{equation}
The first expression is our approximate description of evolution to 
an Eulerian sphere, and the second is simply the spherical collapse 
model. 

Substituting these approximations into equation~(\ref{eqn:fitEC}) 
yields 
\begin{equation}
\frac{R_k^i}{R_2}[1-D(t)\lambda_k] = 
\left\{ \begin{aligned}
 &  \frac{R_k}{R_2}-1 + \frac{1-D(t)\delta_i/3}{a_e(t)/a(t)}, \\
 &  \frac{R_k}{R_2}-1 + \frac{1-D(t)\lambda_2}{a_e(t)/a(t)},
\end{aligned}\right.
\end{equation}
for which the associated overdensities are  
\begin{equation}
\begin{split}
\rho  = & \prod_{j=1}^3 \frac{R_j^i}{R_j} \\
=  & \left\{\begin{aligned}
& \prod_{j=1}^3 \frac{a(t)}{a_e(t)}\frac{1-D(t)\delta_i/3}{1-D(t)\lambda_j}\left(\frac{R_2}{R_j} + \frac{a_e(t)}{a(t)}\frac{1-R_2/R_j}{1-D(t)\delta_i/3}\right) ,\\
& \prod_{j=1}^3 \frac{a(t)}{a_e(t)}\frac{1-D(t)\lambda_2}{1-D(t)\lambda_j}\left(\frac{R_2}{R_j} + \frac{a_e(t)}{a(t)}\frac{1-R_2/R_j}{1-D(t)\lambda_2}\right), 
\end{aligned}\right.
\end{split}
\label{eqn:ECrhoec}
\end{equation}
where $a_e(t)$ is approximated by equation~\eqref{eqn:ae}.
Compared to equation~\eqref{eqn:ECrhosc} the last term in the first 
approximation is the extra factor 
accounting for the non-spherical shape in the final volume. 
For $R_1=R_2=R_3$, this extra factor is unity and equation~\eqref{eqn:ECrhoec} 
has the same expression as in the spherical case.

\section{Perturbative treatment of redshift space distortions}
\citet{kaiser} derives a simple relation between the real and 
redshift space power spectra.  \citet{fisher95} provides a very 
different derivation of Kaiser's expression.  The Zeldovich 
approximation provides yet another derivation of this relation 
\citep{ecpdf}.  We show that our approximation for ellipsoidal
collapse also reproduces this expression, and then use it to 
derive the next order corrections.  This analysis is useful 
because \citet{rs04} has shown that Kaiser's approximation 
is accurate only on rather large scales.  

Before we present the algebra, it is worth noting that the 
approximation used in the main text for the real space evolution of 
a collapsing ellipsoid follows from 
(i) writing the full ellipsoidal collapse as the Zeldovich approximation 
    times a correction factor which is the ratio of the evolution of 
    a sphere in the Zeldovich approximation to that in the spherical 
    collapse model; and 
(ii) using our convenient approximation to the exact evolution in the 
     spherical collapse model.  
In the perturbative analysis which follows, we have chosen notation 
which illustrates what happens if we keep approximation (i), but 
treat the spherical collapse model exactly rather than approximately.  

We begin by writing the relation between the linear and nonlinear 
overdensities in the spherical evolution model as 
\begin{equation}
 1+\delta \equiv \left(\frac{a(t)}{a_e(t)}\right)^3 = 
  1 + \sum_{n=1}^{\infty}\frac{\nu_n}{n!}\,\delta_l^n.
\end{equation}
Then the exact result has 
\begin{align}
 \nu_2 & =  \frac{2}{3}\left(2 - \frac{g_2}{g_1^2}\right) \nonumber \\
 \nu_3 & =  \frac{2}{9}\left(10-12\frac{g_2}{g_1^2}+\frac{g_{3a}}{g_1^3}
             + 6\frac{g_{3b}}{g_1^3}\right),
\end{align}
where $g_i$ is the $i$th order growth factor
(e.g. Fosalba \& Gazta\~naga 1998).  
For $\Lambda \neq 0$
 $(g_2/g_1^2) \approx - (3/7)\,\Omega^{-1/143}$,  
 $(g_{3a}/g_1^3) \approx - (1/3)\,\Omega^{-4/275}$ and  
 $(g_{3b}/g_1^3) \approx (10/21)\,\Omega^{-269/17875}$
\cite{bchj95}.  
The dependence on $\Omega$ is so weak that we can use the $\Omega=1$ 
values to find $\nu_2\approx 1.62$ and $\nu_3\approx 3.926$.  
In contrast, our previous approximation has 
\begin{equation}
 \nu_n\approx \prod_{i=0}^{n-1} \frac{\delta_{\rm c}+i}{\delta_{\rm c}}
\end{equation}
so $\nu_2 \approx 1.6$ and $\nu_3 \approx 3.52$.

With this notation, our approximation for the evolution of an 
ellipsoid is 
\begin{align}
1+\delta_r &=  \left(1-\frac{\delta_l}{3}\right)^3
          \left(1+\sum_{n=1}^{\infty}\frac{\nu_n}{n!}\delta_l^n\right) 
           \prod_{k=1}^3 (1-\lambda_k)^{-1} \nonumber \\
& = 1 + \grave{\delta_r}^{(1)} + \grave{\delta_r}^{(2)}+\grave{\delta_r}^{(3)}
     + \dots,
\end{align}
where
\begin{align}
\grave{\delta_r}^{(1)}  = & \sum_{j=1}^{3} \lambda_j  \nonumber \\
\grave{\delta_r}^{(2)}  = & \frac{\nu_2}{2}\,\delta_l^2 
                            + \frac{\delta_l^2}{3} 
                            - \sum_{j \neq k} \lambda_j\lambda_k \nonumber \\
\grave{\delta_r}^{(3)}  = & \frac{\nu_3}{6}\,\delta_l^3
   + \frac{17}{27}\,\delta_l^3
   - 2\delta_l\,\sum_{j\ne k} \lambda_j\lambda_k 
   + \lambda_1\lambda_2\lambda_3.
\end{align}
In each expression above, the first terms are those associated with 
the spherical model, and the others are the corrections which 
come from the ellipsoidal collapse.  The terms associated with a 
perturbation theory analysis of the exact ellipsoidal collapse model 
are given by Ohta et al. (2004).  The expressions above show the terms 
which are associated with our simple approximation to the exact evolution.  
They differ slightly from those associated with the exact analysis; 
e.g., the second order expression associated with the exact analysis 
is $(4/17)(\delta_l^2 - 3I_2)$, but they otherwise have the same 
form, suggesting that our simple expression captures the essence of 
the ellipsoidal collapse evolution.

The overdensity in redshift space is 
\begin{align}
1 + \delta_s \equiv& \frac{1 + \delta_r}{1 - \sum_{k=1}^3\grave{g}_k^z\,e_k^2}
             \nonumber \\
= & 1 + \grave{\delta}_s^{(1)}+\grave{\delta}_s^{(2)} +\grave{\delta}_s^{(3)}
    + \dots,
\end{align}
where 
\begin{align}
\grave{g}^z_k  = & \frac{\left\{R_k^i\lambda_k - A_h^i\,
 \left[f_1 \delta_l-
 \left(1+\sum_{n=1}^{\infty}\frac{\nu_n}{n!}\delta_l^n\right)^{-4/3}
 \left(\sum_{m=1}^{\infty}f_m\frac{\nu_m}{m!}\delta_l^m\right)\right]/3\right\}}
{R_k^i(1-\lambda_k) - A_h^i[1 - \delta_l/3 - (1 + \sum_{n=1}^{\infty}\frac{\nu_n}{n!}\delta_l^n)^{-1/3}]} \nonumber \\
 = & f_1\lambda_k + \frac{1}{3}(\frac{\nu_2}{2}f_2-\frac{4}{3}f_1)\delta_l^2
   + f_1\lambda_k^2 \nonumber \\
& +\frac{\delta_l^2}{3}\left[(\frac{\nu_2}{2}f_2-\frac{4}{3}f_1)\lambda_k
  + (\frac{\nu_3}{6}f_3-\frac{2\nu_2}{3}f_1-\frac{5\nu_2}{6}f_2+2f_1)
                           \delta_l\right] \nonumber \\
& + \frac{1}{3}(\frac{\nu_2}{2}f_2-\frac{4}{3}f_1)\lambda_k\delta_l^2
 +f_1\lambda_k\left[\lambda_k^2+\delta_l^2(\frac{\nu_2}{6}-\frac{2}{9})\right]
 + \dots,
\end{align}
and $f_i = {\rm d} \ln g_i/{\rm d} a$.  Thus, 
\begin{align}
 \delta_s^{(1)} = &  \delta_r^{(1)} + \Delta_z^{(1)} \nonumber \\
 \delta_s^{(2)} = & \delta_r^{(2)} + \Delta_z^{(2)} 
                   + \delta_r^{(1)}\,\Delta_z^{(1)} \nonumber \\
 \delta_s^{(3)} = & \delta_r^{(3)} + \Delta_z^{(3)}  
                   + \delta_r^{(2)}\,\Delta_z^{(1)} 
                   + \delta_r^{(1)}\,\Delta_z^{(2)}, 
\end{align}
and
\begin{align}
 \Delta_z^{(1)} = & f_1 \sum_{k=1}^3 \lambda_k\, e_k^2 \nonumber \\
 \Delta_z^{(2)} = & f_1 \sum_{k=1}^3
    \left[\frac{\nu_2}{2}\,\frac{f_2}{f_1}-\frac{4}{3}\right]
      \frac{\delta_l^2}{3}\,e_k^2 \nonumber\\
 &  + f_1^2\,\sum_{k=1}^3 \lambda_k^2\,e_k^2 
    + f_1^2 \sum_{j,k=1}^3 \lambda_j\lambda_k\, e_j^2e_k^2 \nonumber \\
 \Delta_z^{(3)} = & f_1 \sum_{k=1}^3 
  \frac{\delta_l^2}{3}\left[(\frac{\nu_2}{2}\,\frac{f_2}{f_1} 
                             - \frac{4}{3})\,\lambda_k
  + (\frac{\nu_3}{6}\,\frac{f_3}{f_1} - \frac{2\nu_2}{3} - \frac{5\nu_2}{6}\,\frac{f_2}{f_1} + 2)\,\delta_l\right]e_k^2 \nonumber \\
& + 2f_1^2\sum_{j,k=1}^3\,\lambda_j\,\left[(\frac{\nu_2}{2}\,\frac{f_2}{f_1}-\frac{4}{3})\,\frac{\delta_l^2}{3} + \lambda_k^2\right]\,e_j^2e_k^2 \nonumber \\
& + f_1^3 \sum_{i,j,k=1}^3 \lambda_i\lambda_j\lambda_k\,e_i^2 e_j^2 e_k^2.
\end{align}

Our leading order term, $\delta_s^{(1)}$ is the same as that of 
Ohta et al. (2004), who showed that 
 $\sigma_s^2 \equiv \langle(\delta_s^{(1)})^2\rangle$, 
where the angle brackets denote the result of averaging over 
the Euler angles as well as the distribution of the $\lambda_i$, 
leads to Kaiser's formula.  
The logic which leads to our second order term is consistent 
with that of Watts \& Taylor (2001), except that because our 
collapse model considers motions relative to the center of 
the object, it has no term which accounts for the motion of 
the center of mass.  There are additional small differences which 
arise from the fact that our description of ellipsoidal collapse 
is based on our approximate model - had we computed our perturbation 
series expansion using the exact collapse model, we would have 
reproduced their expressions.  

\begin{figure}
\centering
\includegraphics[width=\linewidth]{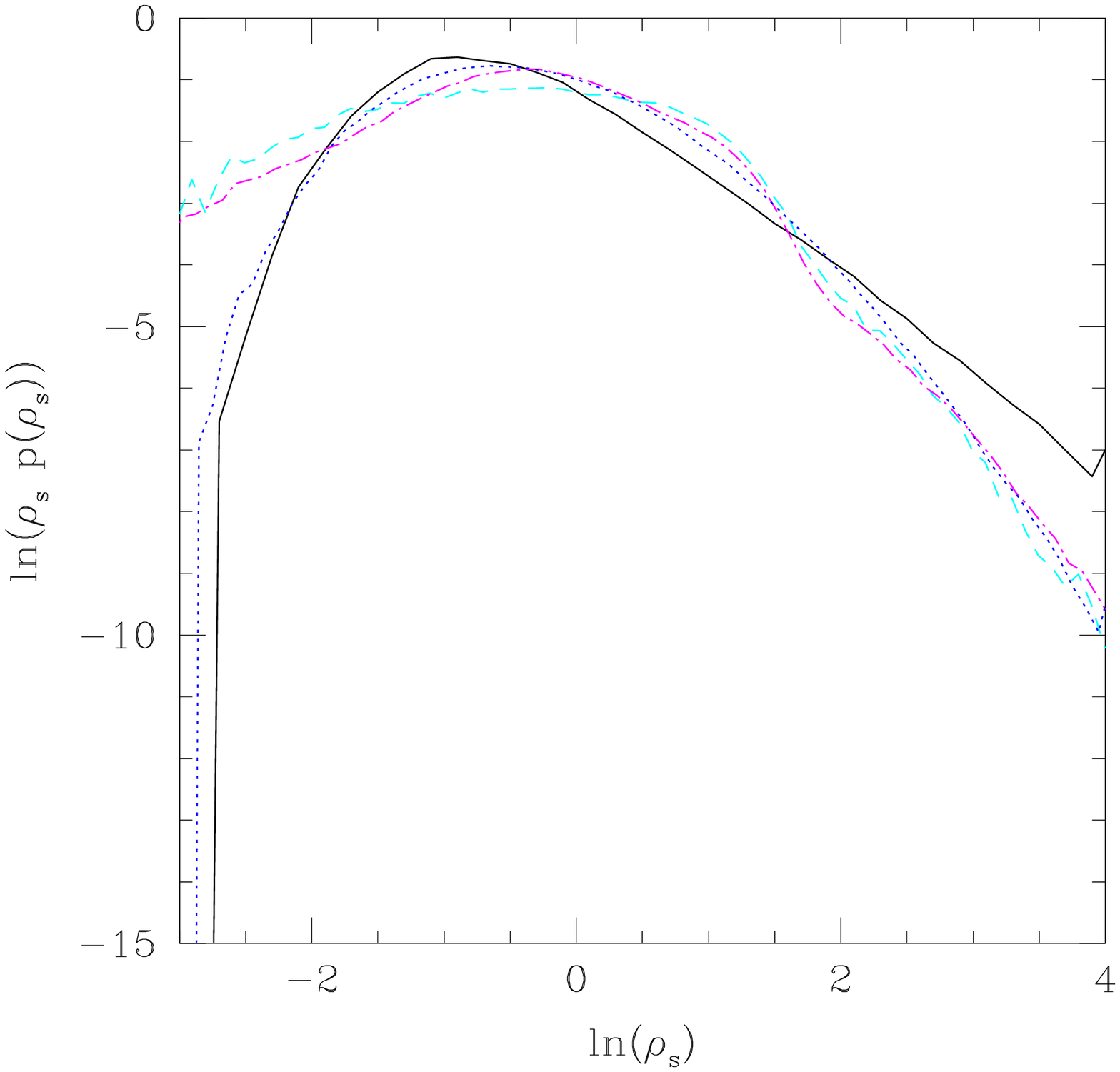}
\caption{Four approximations to the redshift space PDF.  
         Solid line shows the prediction associated with the Zeldovich 
         approximation, but assuming that the associated $\sigma$ 
         is the same for all cells of volume $V$ (Hui et al. 2000).  
         Dashed and dot-dashed curves show the first- and third-order 
         exact ellipsoidal collapse based analyses; these do account 
         for the fact that $\sigma$ depends on the Lagrangian 
         size and shape (e.g. Ohta et al. 2004).  Dotted curve shows 
         the PDF associated with our approximation to the exact 
         evolution.  }
\label{fig:PDFzapprox}
\end{figure}

Figure~\ref{fig:PDFzapprox} illustrates one reason why we have 
used our approximation for the nonlinear dynamics, rather than 
worked with exact perturbative expansions to higher and higher 
orders.  The dotted line shows the redshift space PDF associated 
with our approximation; the main text shows that it provides an 
good description of the simulations.  
The solid line shows the result of using the Zeldovich 
approximation for the evolution, and setting
 $\sigma = \sigma(\bar\rho V)$ 
whatever the shape and overdensity of the initial object \cite{hks00}.  
This results in a PDF with a tail that scales as $\rho_s^{-3}$ 
at high densities --- a signature that the approach has broken down.  
The dashed curve shows the result of using 
equation~(\ref{eqn:ellvar}) for $\sigma$ instead.  This helps 
somewhat --- although the onset of the $\rho_s^{-3}$ tail is shifted 
to larger $\rho_s$, the shape of the PDF in underdense regions is 
adversely affected.  The dot-dashed curve shows the result of going 
to third-order in the exact ellipsoidal collapse model (recall that, 
in this context, the Zeldovich approximation is like the first order 
term).  This modifies the shape of the PDF slightly, but the problem 
at small $\rho_s$ remains.  

\label{lastpage}
\end{document}